\documentclass{emulateapj}
\usepackage{color}

\newcommand{\msun}{$\mathrm{M_{\odot}}$}
\newcommand{\rsun}{$\mathrm{R_{\odot}}$}
\newcommand{\kms}{$\mathrm{km\,s^{-1}}$}

\newcommand{\drvm}{$\Delta \mathrm{RV_{max}}$}

\newcommand{\mym}{$\mathrm{mergers\,yr^{-1}\,M_{\odot}^{-1}}$}
\newcommand{\fb}{$f_{\rm bin}$}

\slugcomment{Draft Version \today}

\shorttitle{Characterizing the WD Binary Population}
\shortauthors{Maoz, Badenes \& Bickerton}

\begin{document}

\title{Characterizing the Galactic White Dwarf Binary Population 
with Sparsely Sampled Radial Velocity Data}

\author{Dan Maoz \altaffilmark{1}, Carles Badenes\altaffilmark{2,1,3}, Steven J. Bickerton \altaffilmark{4}}

\altaffiltext{1}{School of Physics and Astronomy, Tel-Aviv University, Tel-Aviv 69978, Israel; maoz@astro.tau.ac.il}

\altaffiltext{2}{Department of Physics and Astronomy and Pittsburgh Particle Physics, Astrophysics, and Cosmology Center
  (PITT-PACC), University of Pittsburgh, 3941 O'Hara Street, Pittsburgh, PA 15260, USA; badenes@pitt.edu}

\altaffiltext{3}{Benoziyo Center for Astrophysics, Weizmann Institure of Science, Rehovot 76100, Israel}

\altaffiltext{4}{Department of Astrophysical Sciences, Princeton University, Peyton Hall, Ivy Lane, Princeton, NJ
  08544-1001, USA; bick@astro.princeton.edu}

\begin{abstract}
  We present a method to characterize statistically the parameters of a detached binary sample -- binary fraction, separation
  distribution, and mass ratio distribution -- using noisy radial-velocity data with as few as two, randomly spaced, epochs per
  object.  To do this, we analyze the distribution of \drvm, the maximum radial-velocity difference between any two epochs for the
  same object.  At low values, the core of this distribution is dominated by measurement errors, but for large enough samples
  there is a high-velocity tail that can effectively constrain the parameters of the binary population.  We discuss our approach
  for the case of a population of detached white-dwarf (WD) binaries with separations that are decaying via gravitational wave
  emission.  We derive analytic expressions for the present-day distribution of separations, integrated over the star-formation
  history of the Galaxy, for parametrized initial WD separation distributions at the end of the common-envelope phase. We use
  Monte Carlo techniques to produce grids of simulated \drvm\ distributions with specific binary population parameters, and the
  same sampling cadences and radial velocity errors as the observations, and we compare them to the real \drvm\ distribution to
  constrain the properties of the binary population. We illustrate the sensitivity of the method to both the model and
  observational parameters. In the particular case of binary white dwarfs, every model population predicts a merger rate per star
  which can easily be compared to specific type-Ia supernova rates. In a companion paper, we apply the method to a sample of $\sim
  4000$ WDs from the Sloan Digital Sky Survey.  The binary fractions and separation distribution parameters allowed by the data
  indicate a rate of WD-WD mergers per unit stellar mass in the Galactic disk, $\sim 1\times10^{-13}$ \mym, remarkably similar to
  the rate per unit mass of Type-Ia supernovae in Milky-Way-like galaxies.
\end{abstract}


\section{INTRODUCTION}
\label{sec:Intro}

Stellar multiplicity is a fundamental piece in our current picture of stellar formation and evolution. Modern studies of stellar
multiplicity aim to constrain the fraction of stars with companions (or multiplicity fraction, $f_{\rm m}$), the distribution of
separations, and the dependence of these parameters on variables like stellar mass, age, and metallicity. Different observational
techniques are used to probe different separation and flux contrast regimes
\citep{DuquennoyA.1991,Makarov2005,Mason2009,Metchev2009,Raghavan2010}. Short-period binary systems are of particular interest as
the ancestors of future interacting binaries, from cataclysmic variables to novae, high and low-mass X-ray binaries, supersoft
X-ray sources, and Type Ia supernovae. However, the fundamental properties of short-period binaries in the Galaxy are still poorly
constrained. This has important implications for testing specific scenarios for the formation of multiple stellar systems
\citep{Tohline2002,Goodwin2005,Bressert2010,Marks2011b}, stellar population synthesis models
\citep{bruzual93:isochrones,Conroy2009,Marks2011a}, and birth rate calculations for their interacting descendants
\citep{Belczynski2004,ruiter09:SNIa_rates_delay_times}.

The first modern spectroscopic survey for close stellar binaries was performed by \cite{DuquennoyA.1991}, who examined 164 F and
G-type stars, taking more than 4200 individual radial velocity (RV) measurements.  Such surveys are extremely labor intensive,
because they need to obtain enough RV measurements of each target to either confidently discard multiplicity up to a certain
period threshold, or to derive an orbital solution of sufficient quality. For example, the recent study by \citet{Raghavan2010}
examined 454 solar-type stars with different techniques (including RVs), only a factor 3 improvement over \cite{DuquennoyA.1991}
in almost twenty years. The advent of large spectroscopic data bases like the Sloan Digital Sky Survey
\citep[SDSS,][]{york00:SDSS_Technical} opens the possibility to take RV surveys to the next level, allowing millions of RV
measurements of hundreds of thousands of different stars with well-calibrated and stable instrumental set-ups.

In this paper, we present a method to characterize binary populations statistically based on large samples of noisy RVs, with only
a few epochs per target, but without followup observations, confirmations of real binaries, and derivations of orbital parameters
for individual systems. The observable that we analyze is the distribution of maximum RV differences, \drvm, which is
straightforward to obtain from a set of RV measurements for a sample of stars. This distribution contains information about the
orbital velocity amplitudes of some members of some of the close binaries in a sample of stars.  The use of velocity differences
assures that the (uninteresting in the present context) systemic velocities of the stars are subtracted out of the dataset, along
with any other time-constant velocity offsets (wavelength calibration errors, gravitational redshifts, etc.). Naturally, the more
times a system is observed, the higher the chances of catching its full orbital RV variation amplitude, but even with only two
epochs per system, some fraction of that amplitude will be probed. Apart from the observed RVs themselves and their sampling
times, accurate knowledge of the distribution of RV measurement errors is essential. We illustrate how the sensitivity of the
method depends on these observational characteristics.

\cite{Tutukov1986} outlined such an approach schematically, as a way to discover close binary white dwarfs (WDs).
The first application of this kind of technique (though in a more rudimentary way) was by \cite{Maxted2005}. A slightly different
approach that has been applied to SDSS data is described in \cite{ClarkBenjaminM.2011}.  Here, we present the method in the
context of the problem of characterizing a population of detached WD binaries whose separations are decaying via
gravitational wave emission.  In a companion paper (Badenes \& Maoz 2012; Paper II), we apply the method to a sample of $\sim
4000$ WDs from the SDSS with multiple-epoch spectra, we constrain the sample's binary population parameters, and we estimate its
gravitational-wave-driven merger rate.

\section{MONTE CARLO SIMULATION OF THE \drvm DISTRIBUTION OF A BINARY POPULATION}
\label{sec:monte-carlo-simul}

We now simulate the \drvm distribution of an assumed binary population. Our methodology can be applied to any binary
population, but we will focus here on detached binary WDs in the Galactic disk, with the aim, in Paper II, of finding
the regions of parameter space that describe a binary population that are allowed by the observed \drvm\ distribution for a sample of SDSS WDs.  We
simulate WD binaries with properties drawn, in a Monte-Carlo process, from possible families of distributions of these
properties, as detailed below. We then ``observe'' the simulated systems with the sampling sequences and the velocity error
distributions of the real data, to produce each model \drvm distribution.  A detached WD binary will merge, due to loss
of energy and angular momentum to gravitational wave emission, within a time dictated by its separation and its
component masses.  A corollary of every population model will therefore be a prediction of the model's WD merger rate
(whether sub-Chandrasekhar or super-Chandrasekhar), which can be directly compared to observed type-Ia supernova rates, or to the
rates of other transient events that are candidates for the outcomes of such mergers.  We describe below each step in
this modeling process.

Our modeling approach is distinct from that of ``binary population synthesis'' (BPS) calculations such as \cite{Toonen2011},
\cite{Mennekens2010}, \cite{Wang2010}, or \cite{ruiter09:SNIa_rates_delay_times}. In BPS, one begins by simulating a population of
main-sequence binaries, with a chosen mass and separation distribution, and one then attempts to follow the complex stages of
stellar and binary evolution of each system, including mass transfer, mass loss, common envelope evolution, and so on. BPS
calculations have many free parameters. Apart from the parameters that specify the initial conditions, there is a variety of
parametrized ways to approximate the physics of various stages of evolution, particularly the common-envelope phases. Because of
this variety and freedom, there is a great range among the predictions of different BPS calculations for the characteristics of
the final WD populations. Instead, our approach is to parametrize in a simple mathematical way the properties of the binary
population at the {\it end} of its complex physical evolution -- for binary WDs, at the end of the last common-envelope
phase. Beyond that phase, there is only well-understood and easily modeled orbit decay due to gravitational wave emission. The
general forms of our parametrizations for the component masses and separations at this evolved stage are guided by direct
observations, by BPS results, or by educated guesses. They are thus more realistic than those based solely on a particular BPS
realization, but they also allow investigating a larger parameter space for what the binary population might actually be like. The
real RV measurements can then select the particular allowed regions of this parameter space.

\subsection{WD primary mass}
\label{primwdmasses}

For every simulated WD system (either single or binary) we begin by assigning a primary mass (`primary' and `secondary' refers
here to the larger and smaller mass, respectively, not to which star will dominate the spectral energy distribution). We draw the
primary mass, $m_1$, from the observed distribution of WD masses determined by \cite{kepler07:WD_mass_distribution} for 1859 hot
(effective temperature $T_{\rm eff} > 12000$ K) and bright ($g \leq 19$ mag) DA WDs in the DR4 SDSS catalog.  We do not use the
mass functions implied for the cooler WDs in each class, as Kepler et al. (2007) point out and discuss the uncertainty in the
atmospheric modeling of those cool WD, which likely leads to a systematic over-estimate of their masses.  The mass distribution is
composed of four Gaussian components -- a main, narrow, component centered at 0.58 \msun\ with $1\sigma$ width of 0.047 \msun, and
three weaker components centered at lower and higher masses.  The latest version of the SDSS WD catalog, corresponding to DR7, has
over $17000$ entries \citep{Kleinman2009}\footnote{The catalog has not been published yet, but it was kindly made available to us
  by S. Kleinman (private communication). The version that we use here is from July 2010, and it has $17371$ entries.}.  The
\cite{kepler07:WD_mass_distribution} sample is a subset of the DR7 WD sample that we actually analyze in Paper II, and so its mass
distribution is likely quite representative of that of the WDs that we see in our sample.  As a consistency check, we have
measured the mass distribution for hot and bright DA WDs in the DR7 catalog, using the $T_{\rm eff}$ and effective gravity,
$\log{g}$, values fitted by Kleinman et al., and the cooling curves of \cite{Fontaine2001} \footnote{We obtained an updated
  version of these cooling curves from \url{http://www.astro.umontreal.ca/~bergeron/CoolingModels/}.}. We have confirmed with a KS
test that this distribution is very close to the one obtained by \citeauthor{kepler07:WD_mass_distribution} for the DR4 WDs.

It is likely that, contrary to our assumption above, WD primaries do not have the same mass distribution as single WDs.  Mass
transfer and mass loss in close binaries can lead to either an increase or a decrease in the primary WD's final mass. In the BPS
calculations of \cite{Claeys2011}, based on the code of \cite{Izzard2006}, the primaries in WD pairs with separations of $<14
R_\odot$ have a broad mass distribution, between 0.3 and 0.9 \msun\ (J. Claeys 2011, private communication).  However, we find
that even rather drastic changes in the primary mass function have only a small effect of the \drvm\ distribution and on the
merger rate. We have changed the center of the main Gaussian mass component to as low as 0.4 \msun\ or as high as 0.8 \msun, and find
a negligible effect of the \drvm\ distribution.  Increasing the $1\sigma$ width to 0.2 \msun\ also has only a small effect on
\drvm\ (at the level of, e.g., increasing the distribution by $\sim 20$\% at \drvm=300\kms; see below). Our conclusions will
therefore not depend on the particular form we have assumed for the WD primary mass distribution. The weak dependence of \drvm, in
general, on the details of the binary mass components, is discussed further in Section~\ref{resultssection}.

\subsection{Binary fraction}
\label{sec:binfrac}

One major binary population parameter to be constrained by data is the fraction of objects of a class that is in binary systems.
As we will see below, an analysis of a \drvm\ distribution can be sensitive only to binary systems with velocities in the tail of
the distribution, beyond a ``core'' that is dominated by single systems and random velocity errors. For example, for the SDSS WD
sample analyzed in Paper II, this threshold is at \drvm\ $\gtrsim $250 \kms. For the range of possible binary component masses in
a population, this velocity-difference threshold effectively puts an upper limit on the binary separations that can be probed.  In
an extreme-mass-ratio WD binary with masses $m_1=1.2$ \msun\ and $m_2=0.2$ \msun, the secondary ($m_2$) will achieve such orbital
peak-to-trough velocity amplitude if the separation is $\lesssim$0.05 AU. For individual WDs with low-noise measurements and long
temporal baselines, the SDSS data might be able to detect binaries with larger separations, but our analysis does not single out
such systems. Therefore, we will define the binary population parameter, \fb, as the fraction of all WD systems (both single
systems and binary systems) that are binary systems with separations $a<0.05$~AU.  The length of an individual exposure determines
the minimum separation that we can detect. For the SDSS sample, the individual exposure times are $\sim$15 min. For the lowest WD
masses, $\sim 0.2$ \msun, this corresponds to separations of $\sim 10^5$~km, or only about 10 WD radii. Thus, the SDSS data can
probe binary separations ranging from close to Roche-lobe overflow, and out to 0.05 AU, which corresponds to periods of about 4
days for typical WD masses. Since all double-degenerate (DD) systems that merge within a Hubble time have periods shorter than
$\sim 12$~hours, this is more than adequate to provide an estimate of the local WD merger rate.

However, the binary fraction cannot be considered independent of the masses of the WD components. From the initial-final mass
relation for stars and WDs \citep[e.g.][]{Williams2009}, it is known that WDs with masses less than $m_{\rm lim}\approx
0.45$~\msun\ have not had enough time to form in isolation over the age of the Universe, and therefore such WDs must be in
binaries, with separations that permit interactions in the course of the stellar evolution of the components. This was first shown
directly by \cite{Marsh1995}, who found that five out of the seven WDs that they studied, with masses below $ 0.45$~\msun\, had
WD companions with periods $P < 5$~d. \cite{Rebassa2011} have shown that, in binaries composed of a main-sequence
star and a WD of mass $\lesssim 0.5$~\msun, the period is generally $\lesssim$ a few days.  Most recently, \cite{Brown2012}
mined the SDSS photometrically for WDs with masses $<0.25$~\msun. After follow-up spectroscopy, they found that such low-mass WDs
are always, or almost always, in close binaries, with periods of less than 1 day, and often with a relatively massive WD (or
possibly neutron star) companion.  The extremely low-mass WDs actually found and followed up by \cite{Brown2012} turn out to
have even lower masses, always $<0.20$~\msun. It is not known at what precise mass the {\it close} ($P\lesssim 1$~d) binary
fraction becomes 100\%.

\begin{figure}

  \centering
 
  \includegraphics[angle=90,scale=0.9]{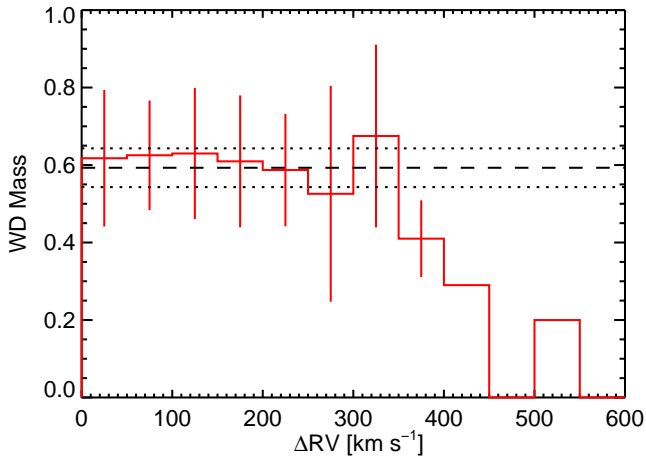}

  \caption{Mean and rms (shown with error bars) masses of the photometric primaries in the observed sample of Paper~II, based on
    spectral modeling of the WDs, for various bins of \drvm. The WDs with large \drvm ($> 200$~km~s$^{-1}$) which reflect the
    close binaries in the sample, have masses $\sim 0.6$~\msun, similar to single WDs, whose mean and rms mass range is shown by
    the horizontal dashed and dotted lines. The bins without error bars are based on only one system each, and the 375 \kms\ bin
    is based on two systems.  }
\label{Mvsdrvm}
\end{figure}

To account for this effect, we adopt a limit $m_{\rm lim}=0.25$~\msun, such that when our simulated primary WD mass is smaller
than $m_{\rm lim}$, we always assign it to be in a binary.  The fraction of the \citeauthor{kepler07:WD_mass_distribution} mass
function that is below this mass means that, effectively, we assign 0.07\% of the WDs in the simulated sample to be in binaries in
which one of the WDs is less massive than $0.25$~\msun. Naturally, close companions likely exist also around all WDs with somewhat
higher masses, e.g., $<0.35$~\msun\ (which constitute 2.5\% of the Kepler et al. 2007 WDs) or $<0.45$~\msun\ (9\%). However, we do
not know for a fact, for those higher masses, that the orbits are within the 0.05~AU limit that we have set above, when defining
close binarity. Indeed, the high-\drvm\ tail of our observed sample in Paper II is {\it not} dominated by low-mass objects, but
rather by typical, $\sim 0.6$~\msun, WDs, as shown in Fig.\ref{Mvsdrvm}.  Thus, \fb\ in our simulations is the fraction of
systems with separations $a<0.05$~AU, and with both components with masses above $m_{\rm lim}=0.25$~\msun. Every simulation
includes an additional population of extremely-low-mass ($m<m_{\rm lim}$) WDs that are all in binaries and are not counted in \fb.
To investigate how our results depend on the choice of $m_{\rm lim}$, we have also calculated models with $m_{\rm
  lim}=0.35$~\msun\ and 0.45\msun. As was the case, above, with variations in the primary WD mass function, we find only weak
effects on the \drvm\ distribution, as long as the total fraction of binaries is equal (e.g., a model with $m_{\rm
  lim}=0.25$~\msun\ and \fb=0.05 gives a very similar \drvm\ distribution as a model with $m_{\rm lim}=0.35$~\msun\ and \fb=0.025,
for which the total binary fraction (including systems with a WD of mass $<m_{\rm lim}$) equals 0.05. Thus, our conclusions will
not depend on the exact choice of $m_{\rm lim}$.

For the fraction \fb\ of the simulated WDs, as well as the extremely-low-mass WD binaries, we assign additional binary
parameters, as described below. To the remaining $1-$\fb\ WDs, we assign an orbital velocity of zero, and skip directly
to the allocation of random velocity errors at several observing epochs (Section~\ref{secvelerr}). The maximum
difference between these random errors for each such non-close-binary WD will then constitute the simulated \drvm\ for
that WD.

\subsection{WD secondary mass}
\label{secwdmasses}

The mass of the secondary, $m_2$, is not likely to be drawn from the same distribution as $m_1$.  Already in main-sequence
binaries, it is clear that the secondary mass is not drawn from the initial mass function, but rather from a mass-ratio
distribution that is approximately flat \citep{Raghavan2010}.  However, there is little in the way of observational or theoretical
guidance for choosing the mass distribution of post-common-envelope WD secondaries.  Of the $\sim$40 known DD systems, only 10 are
double-lined, i.e., with detected spectral features from both components. For these systems, both WD masses can be determined, but
the number is still too small to say much about the mass distribution. In the BPS calculations of \cite{Claeys2011} the
secondary WDs have a roughly flat mass-ratio distribution.  We therefore choose the following parametrization.  In cases where
$m_1$ is above $m_{\rm lim}$, if the simulated system is a binary, we draw the secondary WD mass from a power-law distribution in
mass ratio,
\begin{equation}
P(q)\propto q^\beta ,~~~~
q\equiv m_2/m_1 ,
\end{equation}
with $m_2$ between $m_{\rm lim}$ and $m_1$.  The power-law index $\beta$ is one of the parameters that we vary among the
realizations of our simulation, in order to constrain the properties of the WD binary population. In cases where the primary in
the Monte-Carlo draw was below $m_{\rm lim}$ (and therefore the star is always in a binary), the second star is chosen with equal
probability between $0.2$~\msun\ and $1.2$~\msun. In this scheme, since the typical mass primary has a mass $\sim 0.6$~\msun, the
secondary will have, on average, a mass of $\sim 0.4$~\msun, reflecting the observed commonness of low-mass WDs in close binaries
\citep[e.g.][]{Rebassa2011}.

\begin{figure}

  \centering
 
  \includegraphics[angle=90,scale=0.9]{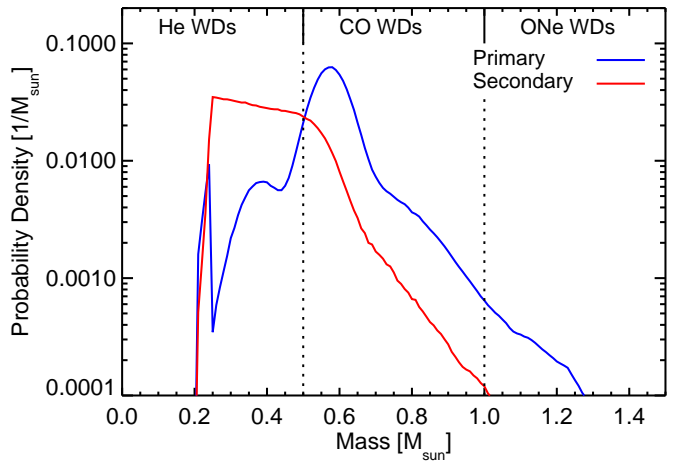}

  \caption{Probability density distribution of primary (blue) and secondary (red) masses in our model binary WD population. A
    binary fraction of \fb=0.05 is assumed in this example, and this sets the height of the extremely low-mass WD peak
    ($<0.25$~\msun), in which WDs are always in close binaries. A flat mass-ratio distribution ($\beta=0$) is also assumed in this
    example.  We note that the two distributions are not independent.  For illustration, the theoretical boundaries between He,
    CO, and ONe WDs in isolated stars are shown with vertical lines. } \label{fig-massDist}
\end{figure}

For illustration purposes, we have plotted these primary and secondary mass distributions in Figure \ref{fig-massDist}, with
$\beta=0$. We have overlaid the standard boundaries between He, CO, and ONe cores in isolated WDs (0.5 and 1.0 \msun,
respectively), but we note that these boundaries might not apply to close binary systems \citep[see][]{PradaMoroni2009}.

\subsection{Separation distribution}
\label{secsepdist}

Next, we assign to each simulated DD system a separation, and hence we need to consider what are the possible separation
distributions for close binary WDs.  The separation distribution at the time the WDs emerge from their final common envelope phase
is unknown observationally, while theoretically it is a longstanding, complex, and unsolved problem \citep[see][for a recent
review]{Ivanova2011}.  The close WDs may undergo either one or two common-envelope phases \citep{Woods2012}, which could,
in principle, lead to a complicated separation distribution. Nevertheless, BPS calculations, as well as some more sophisticated
treatments \citep{Deloye2010}, suggest a power-law separation distribution with a negative index, over the range of separations
that we consider here, $\sim 0.1 - 10$~\rsun.  In the BPS calculations of \cite{Claeys2011}, over the range of separations that we
consider, the post-common-envelope initial WD separations indeed are roughly constant per logarithmic interval (J. Claeys, private
communication). The same is true in the BPS models of \cite{Yungelson2010}, at least for separations above $\sim 1$~\rsun\
(L. Yungelson, private communication).  The $\sim t^{-1}$ Type-Ia supernova delay-time distributions generally predicted by BPS
models for the DD channel would not arise if the WD initial separation distributions were not approximately of the above form
\citep[see, e.g.][]{Maoz2011}.

Therefore, we will assume an initial WD separation distribution that is a power law, with an index that is a free parameter to be
constrained by observations.  Whatever the initial distribution, orbital decay due to gravitational wave emission will immediately
begin to modify it, as all of the orbits shrink and the innermost systems merge. Furthermore, the actual distribution at any
particular time will be the sum of the distributions of many populations of different ages, that have evolved over different
amounts of time.  We now calculate the form of this eveolved, time-integrated, distribution.

\subsubsection{Evolution of a binary separation distribution due to gravitational wave losses}  

\begin{figure*}

  \centering
 
  \includegraphics[angle=90,scale=0.9]{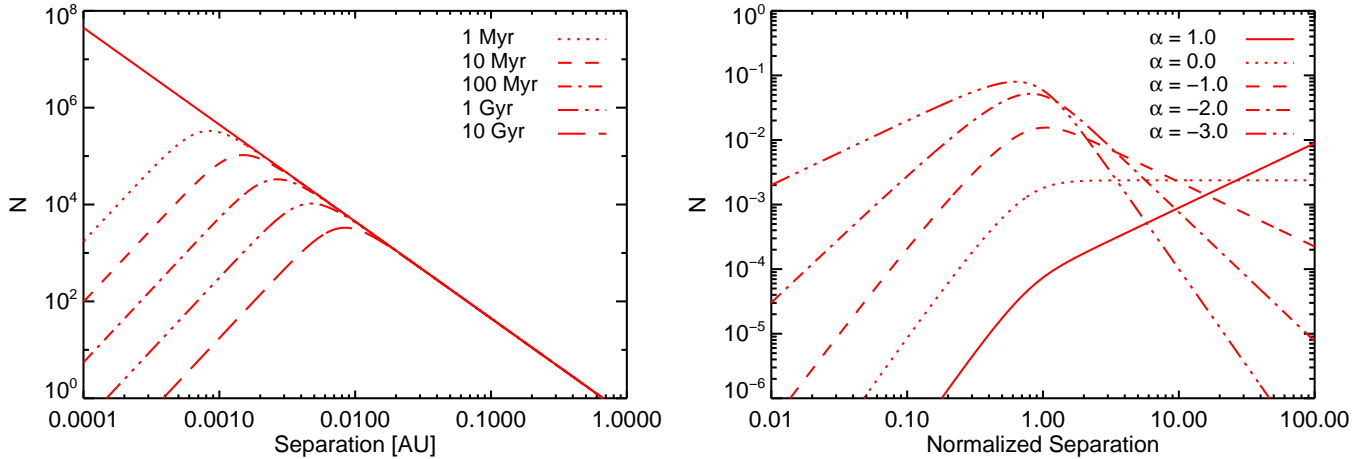}

  \caption{{\it Left}: Example of the gravitational-wave-driven time evolution of the separation distribution of a population with
    an initial power-law separation distribution of index $\alpha=-2.0$ (Eq.~\ref{eqsinglepop}).  In this example, both WDs are of
    $0.55$ \msun. {\it Right}: Separation distributions for various values of $\alpha$, time-integrated over a constant
    star-formation rate over 10~Gyr in the Galactic disk (Eqns. \ref{eqtimeint}-\ref{eqtimeintlog}).  Separations are relative to
    the initial separation of two $0.55$ \msun WDs that will merge within 10~Gyr, $a=0.01$~AU.} \label{fig-WDSepDist}

\end{figure*}

The separation $a$ of two point masses, $m_1$ and $m_2$, in a circular-orbit binary, will shrink over time due to
gravitational wave energy loss as
\begin{equation}
\frac{da}{dt}=-\frac{K}{4 a^{3}},~~~~
K\equiv\frac{256}{5}\frac{G^3}{c^5}m_1 m_2 (m_1+m_2),  
\end{equation}
where $G$ is the gravitational constant and $c$ is the speed of light \citep{Peters1963}.  From integration, the time $t$
for the system to evolve from separation $a'$ to separation $a$ obeys
\begin{equation}
\label{timetomerger}
a'^4-a^4=Kt.
\end{equation}
Suppose a population of WD binaries is formed at a time $t=0$ (this could be, e.g., a population of WD binaries in the
Galaxy that have just emerged from the common envelope phase). The population has an initial distribution of separations
$n'(a')$. For simplicity, we will assume the initial distribution of WD masses is independent of separation. Systems with
separations in the range $a'$ to $a'+da'$ will migrate, after a time $t$, to a bin $a$ to $a+da$ in the evolved
distribution $n(a,t)$.  Conservation of the number of systems (except for those systems that reach $a=0$ and merge) requires that
\begin{equation}
n(a,t) da=n'(a') da' ,
\end{equation}
or
\begin{equation}
n(a,t)=n'(a')\frac{da'}{da}=n'(a')\left(\frac{a}{a'}\right)^3
\end{equation}
$$
=n'[(a^4+Kt)^{1/4}]\frac{a^3}{(a^4+Kt)^{3/4}}.
$$
If the initial distribution is a power law, 
\begin{equation}
n'(a')\propto a'^\alpha,
\end{equation}
then
\begin{equation}
n(a,t)\propto a^3 (a^4+Kt)^{(\alpha-3)/4}.
\label{eqsinglepop}
\end{equation}
The evolved distribution at time $t$ is thus approximately a broken power law. For separations $a\gg(Kt)^{1/4}$ (i.e.,
much larger than those that can merge within time $t$), the distribution will have approximately the original power-law
slope, $n(a)\sim a^\alpha$. At separations $a\ll(Kt)^{1/4}$, on the other hand, $n(a)\sim a^3$ (see left panel of Figure
\ref{fig-WDSepDist}). The merger rate versus time from this single-age population will be controlled by the short separation
pairs for which $a\ll (Kt)^{1/4}$, 
\begin{equation}
\frac{dn}{dt}=\frac{dn}{da}\frac{da}{dt}\propto n(a,t)~ a^{-3}\sim t^{(\alpha-3)/4}.   
 \end{equation}
This is a well-known result for the delay-time distribution of the gravitational-wave-induced 
mergers of a single-age population, having an initial separation distribution that is a power law of index
$\alpha$ \citep[e.g.][]{Greggio2005,Totani2008}.

Let us consider now a series of binary WD populations, each with an initial separation distribution $n'(a')\propto a'^\alpha$,
being produced at a rate $R(t)$ between $t=0$ and the present age of the Galaxy, $t_0$.  The present-day distribution
will be
\begin{equation}
N(a)=\int_0^{t_0} R(t_0-t) n(a,t) dt
\end{equation}
$$
\propto\int_0^{t_0} R(t_0-t)a^3 (a^4+Kt)^{(\alpha-3)/4} dt .
$$
Assuming, for instance, a constant star-formation rate over the age of the Galaxy, then also $R(t)={\rm const.}$ (Even if the
star-formation history is ``bumpy'', the WD formation history will be the convolution of the star-formation history with a broad,
$\sim t^{-0.5}$, kernel, that describes the WD supply [e.g., Pritchet et al. 2008], and which will smooth out the WD production
rate).  The integral then gives
\begin{equation}
N(x)\propto x^{4+\alpha} [(1+x^{-4})^{(\alpha+1)/4}-1],~~~~~ \alpha\ne -1,
\label{eqtimeint}
\end{equation}
or
\begin{equation}
N(x)\propto x^{3} \ln(1+x^{-4}),~~~~~ \alpha= -1,
\label{eqtimeintlog}
\end{equation}
where
\begin{equation}
x\equiv\frac{a}{(K t_0)^{1/4}}
\end{equation}
is the separation in units of the separation of binaries that will merge within the age of the Galaxy.  For example, for
$t_0=10$~Gyr and $m_1=m_2=0.55$ \msun, $x=1$ corresponds to $a_0=0.01$~AU, or $1.5\times 10^6$~km, or about 150 WD radii.
The right panel on Figure \ref{fig-WDSepDist} shows $N(x)$ for various values of $\alpha$. $N(x)$ is, again, approximately
a broken power law, with index $\alpha$ at $x \gg 1$. At $x \ll 1$ the power-law index is 3 for $\alpha\ge -1$, and
$\alpha + 4$ for $\alpha\le -1$. 

\subsubsection{Choice of binary separation}
\label{separationchoice}

We use the functional forms in Eqns.~\ref{eqtimeint}-\ref{eqtimeintlog} to model the possible present-day distributions of DD
separations, for various indices $\alpha$ of the initial power-law distributions at formation.  In realizations of our simulation,
we draw the separation of specific component binary masses from the present-day distribution for a particular value of $\alpha$,
with $a$ between $a_{\rm min} =2\times 10^4$~km (contact) and $a_{\rm max}=0.05$~AU.  For the purpose of producing simulated
radial velocities, binaries with periods $<15$~min are assigned zero orbital velocity, as the SDSS exposure length prevents
detection of velocity differences in such cases.  A practical consideration in this process is the large dynamic range that the
distribution $P(a)$ can assume over this range in $a$. As a result, very few simulated binaries might be assigned separations in
ranges that have low probability, and the model expectation values for the the velocity differences due to those binaries will
have large Poisson errors.  To avoid this, we populate the distribution evenly with simulated systems among four decades of
separation $a$ (i.e., from $a_{\rm min}$ to $a_{\rm max}/1000$, from $a_{\rm max}/1000$ to $a_{\rm max}/100$, etc.).  Each binary
system is given a relative weight according to the integral of the separation distribution over the decade it is in.
  
\subsection{Period, orbital velocities, and merger rates} 

Given two masses and an orbital separation, Kepler's law gives the period
\begin{equation}
\tau=2\pi \left(\frac{a^3}{G(m_1+m_2)}\right)^{1/2},
\end{equation}
and the circular orbital velocities,
\begin{equation}
v_1=\frac{2\pi a}{\tau}\frac{m_2}{m_1+m_2}, ~~~v_2=\frac{2\pi
  a}{\tau}\frac{m_1}{m_1+m_2} .
\end{equation}

We assume circular velocities for simplicity, but also because this is the expectation for close binaries that have likely
undergone circularization by tidal forces and common-envelope evolution. For example, there are essentialy no binaries with
$P<12$~d that have any appreciable eccentricity \citep{Raghavan2010}.  It has been recently suggested \citep{ThompsonToddA.2010}
that there might be a preponderance of triple systems among binary WDs that would induce elliptical orbits via the
\cite{Kozai1962} mechanism, leading to a large population of systems with short merger times. We defer the exploration of this
scenario to future work.  Equation~\ref{timetomerger} with $a=0$ gives the merger time, $t_{\rm merge}$, of the simulated
system. The merger rate per WD in the simulated sample is obtained by noting, for a given set of parameters, the fraction of all
of the systems in the simulation that merge within a set time window, divided by that time. Each system is weighted according to
its decade in separation (see section~\ref{separationchoice}, above).  A separate tally is conducted to calculate the merger rate
of only those binaries whose summed masses exceed the Chandrasekhar mass, which may be of relevance for the type-Ia supernova
rate.  For $\alpha\ge -1$, the merger rate is approximately constant, and therefore the time window for numerically calculating
the rate is arbitrary, as long as it is shorter than $t_0$.  The constancy of the rate can be seen from Eq.~\ref{eqtimeint}, by
noting that mergers come from systems with $x<1$, for which $N(x)\sim x^3$, and the merger rate is
\begin{equation}
\frac{dN}{dt}=\frac{dN}{da}\frac{da}{dt}\propto N(a)~ a^{-3}\sim {\rm const.}  
 \end{equation}
For $\alpha<-1$, the merger rate falls with time, but quite slowly for values of $\alpha$
that are not too steep,
\begin{equation}
\frac{dN}{dt}\sim N(a)~ a^{-3}\sim a^{\alpha+1}\sim t^{(\alpha+1)/4} ,  
 \end{equation}
so an accurate numerical merger rate is still obtained in the above scheme.

The merger rate for a given combination of \fb\ and $\alpha$ can also be roughly estimated analytically. If $\alpha\ge -1$, the
majority of pairs have large separations, with long times until merger. The merger rate is therefore set by the integrated effect
of old systems. All binaries with separations of $a <a_0$ at their formation time will contribute (at a constant rate, as we saw)
to the current merger rate, where $a_0$ is the maximum separation binary that merges within $t_0$.  Every choice of component
masses gives a different value of $a_0$.  However, for the range of possible component masses, $a_0$ is between 0.005 and 0.018 AU
and a value of $a_0=0.01$~AU is typical. The typical time until merger of each system is $t_0$. The merger rate per observed WD is
therefore
\begin{equation}
\frac{1}{N_{\rm wd}}\frac{dN}{dt}\approx \frac{f_{\rm bin}}{(1-f_{\rm bin})t_0}\frac{\int_{a_{\rm min}}^{a_0}n(a) da}{\int_{a_{\rm min}}^{a_{\rm max}}n(a) da}
\end{equation}
$$
=\frac{f_{\rm bin}}{(1-f_{\rm bin})t_0}\frac{(a_0/a_{\rm min})^{\alpha+1}-1}{(a_{\rm max}/a_{\rm min})^{\alpha+1}-1} ,
$$
for $n(a)=a^\alpha $. If \fb\ is small, and (as in the present case), $a_{\rm max}$ and $a_0$
are both much larger than  $a_{\rm min}$, then the merger rate is roughly
\begin{equation}
\frac{1}{N_{\rm wd}}\frac{dN}{dt}\approx \frac{f_{\rm bin}}{t_0}\left(\frac{a_0}{a_{\rm max}}\right)^{\alpha+1}.
\label{rateeq}
 \end{equation}
This gives values in good agreement with the merger rates from the numerical Monte Carlo calculation.
One can also see from Eq.~\ref{rateeq} that, in a plot of the parameter space of $\alpha$ vs. $\log f_{\rm bin}$, curves of constant 
merger rate will be straight lines with slope of $1/\log(a_0/a_{\rm max})\approx 1.4$, for $a_0=0.01$~AU
and $a_{\rm max}=0.05$~AU (see Paper II). For $\alpha \ll -1$, most WD binaries are formed with small separations
and therefore merge within a time much shorter than $t_0$. In this case, the merger rate will be controlled by the
supply rate of new WDs, which in turn is set by the star-formation rate. 

\subsection{Inclination, photometric primary, temporal sampling, and velocity
  error}
\label{secvelerr}

We next apply observational effects to each simulated binary sytem. First, a line-of-sight inclination $i$ of the
perpendicular to the orbital plane is chosen from a distribution
\begin{equation}
P(i)\propto \sin i ,
\end{equation}
and the line-of-sight  velocity  is reduced by $\sin i$.

\begin{figure}

  \centering
 
  \includegraphics[angle=90,scale=0.9]{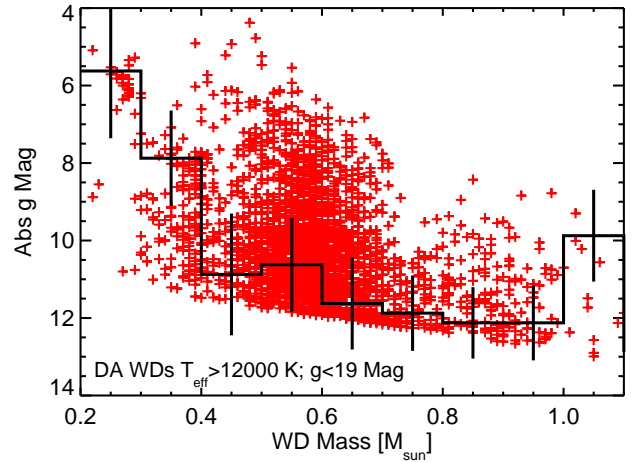}

  \caption{Distribution of absolute $g$ magnitudes as a function of mass for the hot DA WDs in the DR7 catalog. We only
    consider, in this figure, objects with $T_{\rm eff}$ above 12000 K and $g$ mag brighter than 19, for which spectroscopic masses can be
    modeled reliably. The overlaid histogram shows the mode (most frequent $g$ magnitude observed) and the standard
    deviation in bins of 0.1 \msun. There is a large spread in absolute magnitudes at all masses, but low mass WDs
    (below $\sim$ 0.35 \msun) do seem to be intrinsically brighter.} \label{fig-WDMassMag}

\end{figure}

We then need to decide which of the two WDs will be the photometric primary. This could be either the less massive WD, because it
has larger surface area and/or it is younger and hence hotter; or the more massive WD, because it cools more slowly due to its
small surface area and is hence hotter. In practice, these effects compete against each other, and it is difficult to determine
which of the two components will dominate the spectral energy distribution. From an observational point of view, the distribution
of absolute magnitudes as a function of WD mass in the DR7 SDSS catalog shows a large spread about the mean at all masses,
although low mass WDs (below $\sim$0.35 \msun) do seem to be intrinsically brighter (see Figure \ref{fig-WDMassMag}). From a
theoretical point of view, the cooling curves of \cite{Fontaine2001} also predict that, in coeval DD systems, WDs with masses
below $\sim$ 0.35 \msun\ will remain substantially brighter than their more massive counterparts for several hundred Myr (in the
SDSS $g$ filter -- the effect is enhanced in $r$, and diminished in $u$). After about 1 Gyr, the slower cooling of more massive
WDs takes over and makes them brighter, but by this time the predicted magnitudes become fainter than the cutoff in our samples
for most of the volume probed by SDSS. To summarize, it seems reasonable to assume that low-mass WDs, if present, will have a
tendency to dominate the spectral energy distribution of DD systems, but in other cases either of the components may be
dominant. In our Monte Carlo runs, we therefore make the less massive WD the photometric primary when its mass is below 0.35
\msun, but decide randomly between the two WDs when the less massive WD is above this limit.

\begin{figure*}

  \centering
 
  \includegraphics[angle=90,scale=0.9]{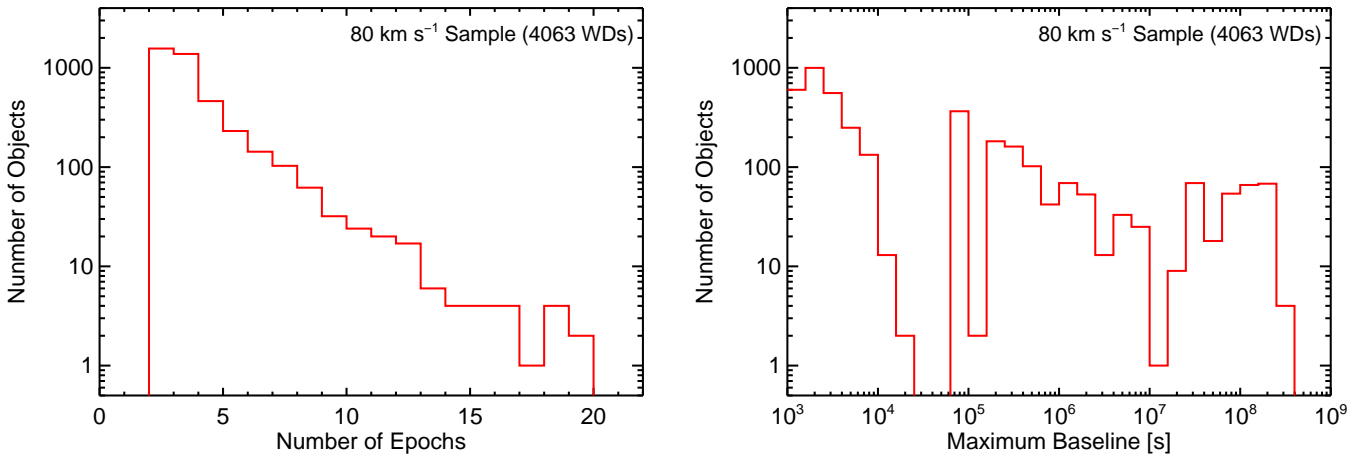}

  \caption{Temporal sampling characteristics for the DR7 WD sample analyzed
in Paper II, and used in the example simulations here. {\it Left:} Distribution 
on number of epochs per object. Most WDs have only two or three epochs, but there 
is a tail of objects with more epochs. {\it Right:} Distribution of maximum
time differences between epochs. A 45~min interval is most common, but longer 
timescales, of order an hour or of several days, are probed in many cases.} 
\label{wdbaselines}

\end{figure*}

Once the photometric primary has been selected, we sample its line-of-sight velocity with the actual distribution of
temporal samplings in the SDSS data.  We do this by choosing at random a particular observation pattern (number of
epochs and time between epochs) from the sample, with a random phase assigned to the first epoch of the sinusoidal
RV curve. Figure~\ref{wdbaselines} shows the distributions of the number of epochs and of the temporal baselines per object for the 
SDSS sample of Paper II.
To each simulated velocity measurement, we add a random error that we draw from a Gaussian distribution, 
with the variance of the Gaussian drawn from the actual distribution of measurement errors for the observed sample  
(see figure~1 in Paper~II for this distribution for the SDSS sample). 
Finally, for every simulated system or non-binary WD,
we find the minimum and maximum observed velocities and calculate \drvm.

For every parameter combination that defines a binary WD population model, we produce $10^5$ WD systems (either single or
binary, according to \fb), and find the fractional prediction for each bin in the model \drvm\ distribution. Multiplied
by the observed WD sample size, this gives the expectation value for that bin. 

\section{Results}
\label{resultssection}

\begin{figure}

  \centering
 
  \includegraphics[angle=90,scale=0.9]{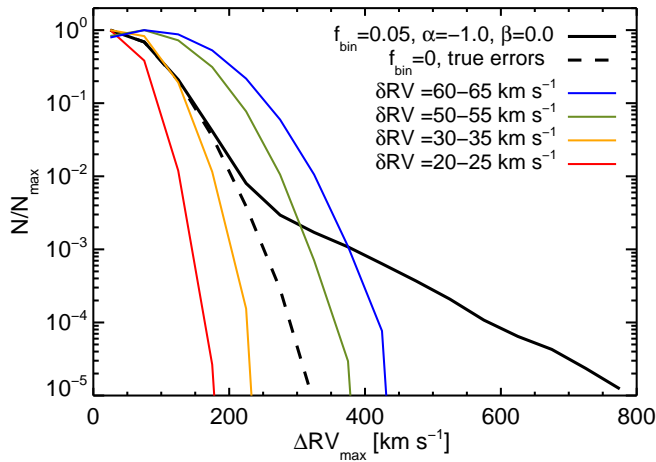}

  \caption{Simulated distributions of \drvm. Solid line shows the distribution
resulting for a WD population with binary fraction \fb=0.05, 
separation distribution 
power-law index $\alpha=-1$, and mass-ratio distribution power-law index
$\beta=0$, when this distribution is observed with the temporal sequences
and RV errors of the DR7 sample presented in Paper~II. The 
dashed line shows the ``core'' of the distribution, obtained by setting 
\fb=0 (no binaries, except for the extremely low-mass WDs, which are always in binaries),
and reflects the part of the full distribution that is due to RV
errors alone. Colored curves show the core distributions obtained when 
assuming incorrect error distributions 
(various narrow distributions of $1\sigma$
error ranges, as marked). Accurate characterization of the errors is thus
essential for characterizing the binary population by means of the 
signal in the tail of the \drvm\ distribution.  
} \label{figerrs}

\end{figure}

Figure~\ref{figerrs} shows the simulated \drvm\ distribution for a model binary population with a particular set of parameters
($f_{\rm bin}$, $\alpha$, $\beta$), when a sample of that population is sampled with a particular set of empirical temporal
sequences and a particular velocity error distribution.  The observational parameters chosen are those of the SDSS sample
presented in Paper~II, and the model parameters are one set among those that reproduce the data.  Along with this distribution, we
plot the predictions for a model with {\it no} binaries in it.  We see that the modeled \drvm\ distribution consists of two
parts. At low \drvm, there is a ``core'' that is dominated by random velocity errors, that have been applied to systems that are
single, or that have low \drvm\ because of one or more causes (low orbital velocity, low inclination, inopportune sampling).
Beyond this core, the distribution has a ``tail'' that reveals those close binaries that happened to have large orbital
velocities, favorable inclinations, and temporal sampling that caught their velocity variations.  For any observational sample,
one can always calculate this zero-binaries core. When compared to the observed \drvm\ distribution, it immediately reveals if and
where in the data there exists a tail of real binary systems. A statistical comparison between the data and a grid of model
distributions can then select the regions of binary population parameter space that are allowed or ruled out by the data.

A reliable estimate of the RV errors is essential, as otherwise underestimated errors can masquerade as a tail, or overestimated
errors can hide a real binary population.  This is also illustrated in Fig.~\ref{figerrs}, where we show several no-binary
predictions that use incorrect error distributions, i.e., error distributions that are different from the one used to generate the
distribution with binaries.

\begin{figure}

  \centering
 
  \includegraphics[angle=90,scale=0.9]{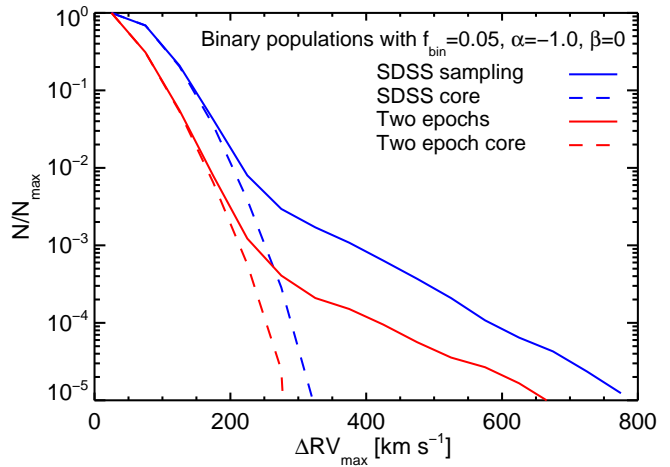}

  \caption{Dependence of \drvm\ distribution on temporal sampling. Blue
curves show the full \drvm\ distribution (solid curves) and the \fb=0 core
(dashed curve), for a chosen set of binary population parameters, and
with the actual DR7 Paper~II sample with its temporal sampling characteristics
(same as Fig.~\ref{figerrs}.) Red curves show the distribution and core
when every object is observed on two epochs only. 
} \label{figsamp}
\end{figure}

The temporal sampling density of the survey will naturally affect both the core and the tail of the \drvm\ distribution.  The more
epochs per object, the greater the chance of catching the full RV variation range of each system.  The core will also grow, but
only as the square root of the number of epochs. Since the fully revealed RV range reaches saturation beyond some number of
epochs, while the core \drvm\ continues to grow as random errors are added in quadrature, there will be a limiting typical number
of epochs, $N$, beyond which the technique is no longer efficient for characterizing the population, and one can, instead, attempt
to fit RV curves to each object.  This will happen when $\sqrt{N}\sigma_v \sim$\drvm$_{+}$, where the latter is the highest value
of \drvm\ for which a sample has some exemplars, and $\sigma_v$ is the typical velocity error.  Fig~\ref{figsamp} illustrates how
the core and the tail of the distribution change when, instead of the full sampling of the SDSS dataset, every object is sampled
only twice.

\begin{figure}

  \centering
 
  \includegraphics[angle=90,scale=0.9]{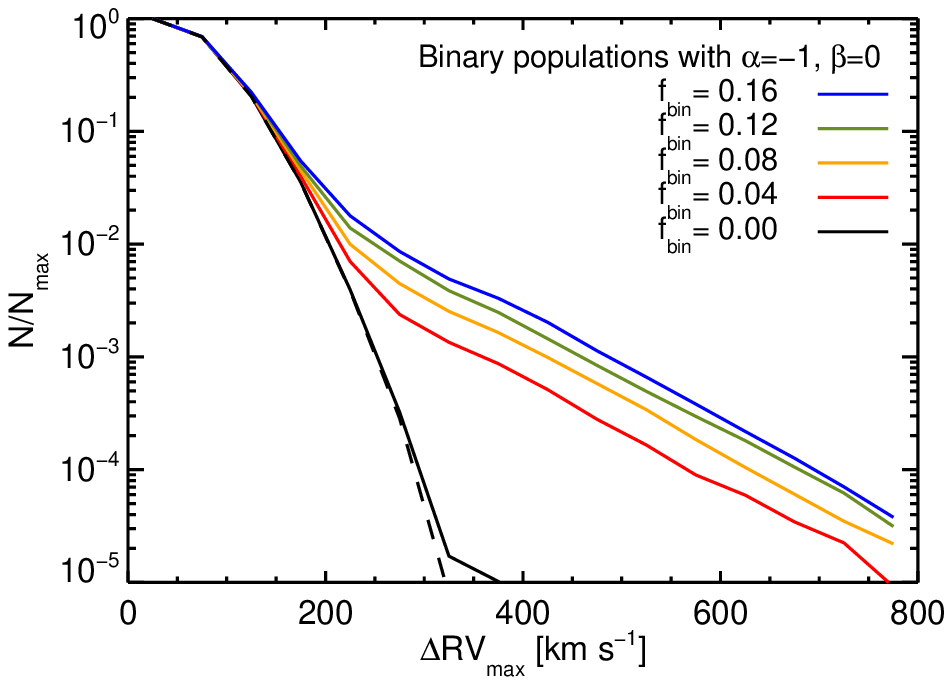}
  \includegraphics[angle=90,scale=0.9]{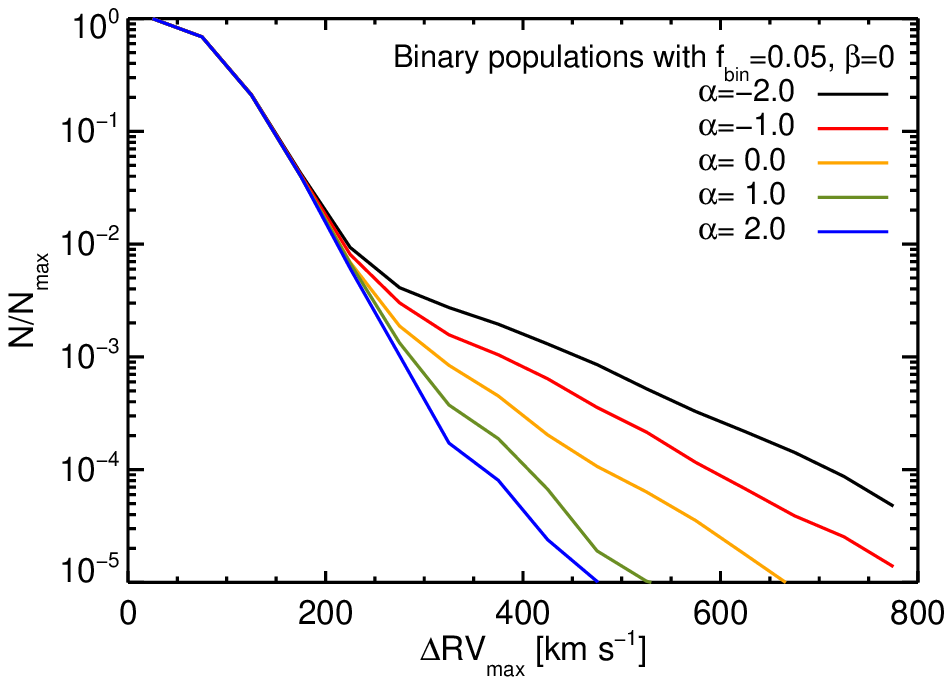}
  \includegraphics[angle=90,scale=0.9]{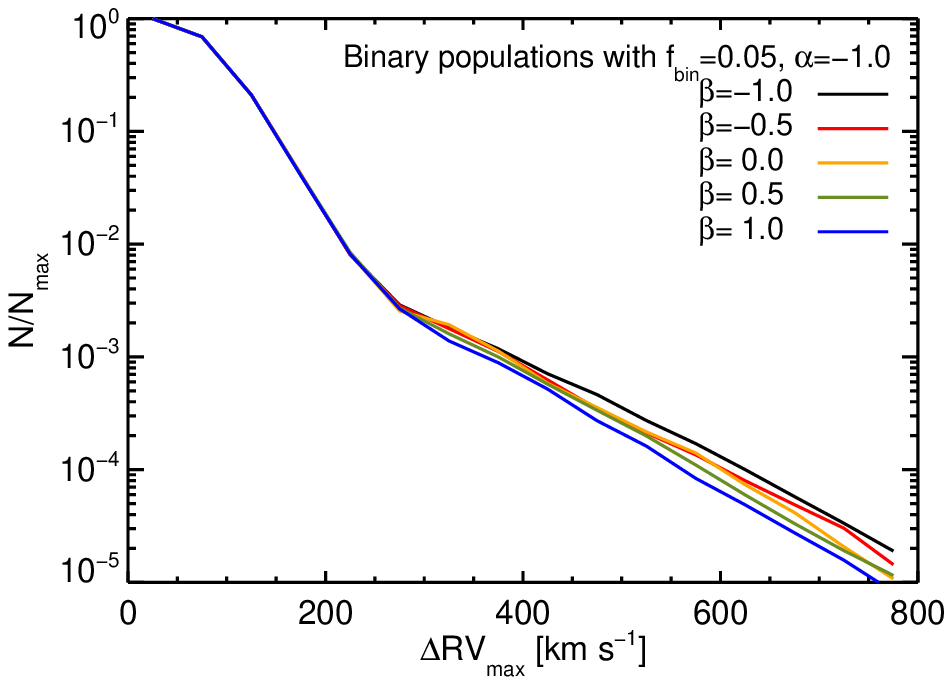}
  \caption{Dependence of the \drvm\ distribution on binary population 
parameters. {\it Top:} Dependence of binary fraction, \fb. The black 
solid curve is for no binaries, except for those that include a $<0.25$~\msun
WD. The dashed curve is with no binaries at all, and shows the bare ``core''
of the distribution, due to the radial velocity errors. {\it Middle:} Dependence
on separation distribution power law index, $\alpha$. {\it Bottom:} The weak 
dependence
on the mass-ratio distribution's power-law index, $\beta$. 
} \label{figfbin}

\end{figure}
Figure~\ref{figfbin} shows how the \drvm distribution depends on the binary population parameters \fb, $\alpha$, and $\beta$.
Qualitatively, increasing \fb\ and decreasing $\alpha$ both have the same effect of increasing the number of small-separation
binaries, and therefore of raising the high \drvm\ tail of the distribution.  Quantitatively, if we were dealing with a population
of binaries with single values of component masses, single values of inclination angle, numerous sampling per object, and no
measurement errors, then the \drvm\ distribution tail would behave as
\begin{equation}
\frac{dN}{dv}=\frac{dN}{da}\frac{da}{dv},
\end{equation}
and the broken power-law separation distribution, $N(a)$, would lead to a broken power-law orbital velocity distribution.
Recalling that, for $\alpha > -1$, the small-separation part of the separation distribution behaves as $N(a)\propto a^3$, and
taking the Keplerian $v\propto a^{-1/2}$, we would then expect the tail to fall steeply as $\sim v^{-9}$ (for all $\alpha>-1$),
with its amplitude depending on $f_{\rm bin}/{(\alpha+1)}$ (the latter factor entering through the normalization of the
distribution).  The $a^\alpha$ branch of $N(a)$ at $a>a_0$ would transform to a $v^{-2\alpha-3}$ power law at velocities $v<v_0$,
where $v_0$ is the orbital velocity of a binary with separation $a_0$. In reality, the distribution of component masses (which
leads to a range of values for $a_0$), the projection of velocities due to inclination, and the sampling (both of which effects
move binaries in the distribution to lower values of \drvm), and the velocity errors (which again mix the distribution), will lead
to a \drvm\ tail with a slope that behaves differently than above, and does depend on $\alpha$.  For $\alpha < -1$, $N(a)\propto
a^{\alpha+4}$, and the \drvm\ distribution tail falls more gently, as $\sim v^{-2(\alpha+11)}$, in the idealized case. Thus, in
principle, $\alpha$ can be discerned in data with low-enough velocity errors, and with large enough samples, such that the slope
of the tail can be measured accurately.

As seen in Fig.~\ref{figfbin}, the \drvm\ distribution is weakly dependent on $\beta$, the power-law index of the binary
mass-ratio distribution.  For a given choice of primary mass, the secondary's velocity will depend on the mass ratio $q$ as
$(1+q)^{-1/2}$.  For the typical $m_1\lesssim 0.75$~\msun\ primary mass, the secondary mass is constrained to the range from $m_1$
down to $m_{\rm lim}=0.25$~\msun, so $q$ is between about 1/3 and 1. Going from high positive values of $\beta$ that favor $q$
near 1, to very negative $\beta$'s that concentrate $q$ for all binaries to be near 1/3, will induce an increase of
$\sqrt{3/2}=1.22$ in the secondary's velocity $v_2$, and an even smaller relative decrease, by 0.93, in $v_1$, if it is the
primary that is selected as the photometric primary. Thus when changing between the extreme values of $\beta$, about half of the
binaries will have their \drvm\ increase by 22\%, and half will decrease by 7\%, or a net increase by 7\% in the \drvm\ of each
bin in the distribution.  Even for the highest velocities that we consider, this is a small change, comparable to the typical
velocity errors.  Furthermore, as we saw above, the \drvm\ distribution is roughly a power law. The transformation $v\rightarrow
v'=kv$, where $k$ is a constant, does not affect the shape of power-law distribution in $v$.

\section{CONCLUSIONS}
\label{sec:conclusions}

We have used Monte Carlo simulations to study how the distribution of maximal radial velocity differences, \drvm, can characterize
a close binary population, in a radial velocity survey where a large number of stars are observed, but with only a few epochs per
star, and with potentially large RV errors. Our focus has been on a survey for double WDs of the kind that we analyze in Paper II,
using the SDSS DR7 WD sample, which has served here as our example survey in terms of observational parameters.  As part of this
modeling process, we required a realistic representation of the present day separation distribution of close WD binaries whose
separations have evolved due to gravitational wave emission. We have therefore derived analytical expressions for the separation
distribution of a population of WD binaries that is continuously formed, its orbits decay, and some of its members merge, over the
lifetime of the Galaxy. With these simulations, we have determined how the \drvm\ distribution depends on the binary population,
which we have characterized using three parameters (describing close-binary fraction, separation distribution, and mass-ratio
distribution), and on the observational parameters of the sample -- RV errors and temporal sampling pattern.

Our main findings are:

\begin{enumerate}
\item The \drvm\ distribution has a core region, that is produced by the random RV errors, and a tail, that can reveal the
  presence of some of the close binaries in a sample. The power of the technique is that, even with large errors and only few (or
  even just two) epochs per object, the close binary population can reveal itself in the tail, provided that the number of objects
  in the sample is large enough. This tail permits statistically constraining the properties of the population, even with sparse
  and noisy data, and without detailed followup work on candidate binary systems.

\item Accurate knowledge of the distribution of radial velocity errors is essential to model properly the \drvm\
  distribution, and thus deduce the contribution of the real binary population to the tail.

\item Steep distributions in initial separation (very negative $\alpha$, with most binaries at small separations) and populations
  with a large close-binary fraction (large \fb) will both result in an increase in the amplitude of the \drvm\ tail. There may
  thus be some degeneracy in the determination of these two parameters.  A change in $\alpha$, however, also changes the shape of
  the tail, and hence, in high-quality datasets (many objects, small errors), these parameters may still be individually
  constrained.

\item The \drvm\ distribution depends weakly on the details of the distributions of the component binary WD masses -- the
  distributions of primary masses and of secondary mass ratios, which in any case are chosen from a a relatively small dynamic
  range.  This binary population characteristic thus cannot be constrained by this kind of survey data. Conversely, not knowing
  the distribution of mass ratios does not affect adversely our ability to constrain the other binary population parameters.

\item The \drvm\ approach allows to estimate the merger rate of a close binary population, based on noisy RV data with few epochs,
  of the kind we consider. This is possible without follow-up observations to obtain full binary parameter solutions for
  candidates, and without necessarily even finding a single binary that will merge within a Hubble time. This ability is a result
  of the statistical nature of our approach.

\end{enumerate}

In Paper~II we apply the methods presented here to the observed SDSS DR7 WD sample, we set constraints on the
local population parameters of close-WD binaries in the Galaxy, and we derive the merger rate of this population, both
in general and for particular component and total mass ranges. We show that the local rate of WD mergers with
super-Chandrasekhar total masses is an order of magnitude lower than the local Type-Ia supernova rate.  The local merger
rate of all WDs, however, is remarkably similar to the Type-Ia suparnova rate.  A large fraction of all WD mergers are
between CO and CO WDs, with total masses not far from the Chandrasekhar limit. If sub-Chandrasekhar mergers result in a
Type-Ia supernova explosion, we may have identified their dominant progenitors.

Apart from Type-Ia supernova explosions, other possible outcomes of WD mergers can be tested with our methodology -- R Corona
Borealis stars \citep[e.g.][]{Longland2011}, or highly magnetic WDs, which have been postulated to result from WD mergers
\citep[e.g.][and references therein]{GarciaBerro2012}. Some $7\pm 3$\% of local WDs have magnetic fields greater than $10^7$~G
\citep{Kawka2007}. Assuming these magnetic fields decay on very long timescales, WD mergers producing all of this population
would need to occur, over 10~Gyr, the age of the Galaxy, at a rate of $\sim (7\pm 3)\times 10^{-12}$~yr$^{-1}$ per WD. In Paper
II, our fit to the observed \drvm\ distribution for the WDs in SDSS indicates a WD merger rate, for total merged masses of
$<1$~\msun, of $1^{+3}_{-0.7}\times 10^{-12}$~yr$^{-1}$ per WD. The rate may thus be sufficient to explain at least some, and
perhaps even all, magnetic WDs with such mergers. 

As another example, about half of WDs with masses below 0.45 \msun\ appear to be single \citep{Maxted2000,
  Napiwotzki2007,Kilic2007}. \cite{Nelemans1998} have suggested formation of such low-mass single WDs via interaction between a
giant star and a close massive-planet or brown-dwarf companion, followed by evaporation or tidal disruption of the
companion. \cite{Kilic2007} have proposed that these WDs have evolved from metal-rich stars whose evolution was truncated by
severe mass loss on the giant branch. Alternatively, \cite{Iben1997} have raised the possibility that the single low-mass WDs are
the merger products of even-lower WDs. Our constraints on merger rates can test this last scenario. In the \cite{kepler07:WD_mass_distribution} WD
mass function, about 8\% of the WDs are in a Gaussian component that
  peaks at around 0.4 \msun. Assuming that half of these WDs
are single, then in the \cite{Iben1997} scenario, about 4\% of the WD population would be the outcome of very-low-mass WD
mergers. This is a similar fraction to the one invoked above in the case of magnetic WDs, and therefore would require a similar
merger rate. Our calculations, however, show that the rate of mergers
  with total masses in the range 0.3--0.5 \msun\ is four
orders of magnitude below the one required by this scenario.  This comes about because each of the merging WDs would need to be
below 0.3 \msun, and such WDs are rare. The longer gravitational orbit decay time at these low masses further lowers the rate. One
could circumvent this argument by invoking large mass loss during the merger process \citep[e.g.][]{Fryer2010}, so that more
massive and common WDs could be involved. However, \cite{Dan2012} find negligible mass loss in their 3D hydrodynamic simulations
of WD mergers.
  
The tools that we have introduced here can also easily be adapted to the characterization of stellar multiplicity based
on large RV surveys in other settings. Examples are ongoing surveys like LAMOST \citep{SuDingQiang1998} and APOGEE
\citep{Prieto2008}, and planned ones, such as BigBOSS \citep{SchlegelD.2011}.

\acknowledgements{We thank Scot Kleinman for making his DR7 WD catalog available to us in advance of publication. We acknowledge
  useful comments and suggestions from 
 Tim Beers, Pierre Bergeron, Warren Brown, Joke Claeys, Brian Metzger, Gijs Nelemans, Marten van
  Kerkwijk, Lev Yungelson, and the anonymous referee. 
We are grateful to all members of the SWARMS team: Mukremin Kilic, Tom Matheson, Fergal Mullally, Tony Piro, Roger
  Romani, and Susan Thompson. DM acknowledges support by a grant from the Israel Science Foundation.

  Funding for the SDSS and SDSS-II has been provided by the Alfred P. Sloan Foundation, the Participating Institutions, the
  National Science Foundation, the U.S. Department of Energy, the National Aeronautics and Space Administration, the Japanese
  Monbukagakusho, the Max Planck Society, and the Higher Education Funding Council for England. The SDSS Web Site is
  http://www.sdss.org/.  The SDSS is managed by the Astrophysical Research Consortium for the Participating Institutions. The
  Participating Institutions are the American Museum of Natural History, Astrophysical Institute Potsdam, University of Basel,
  University of Cambridge, Case Western Reserve University, University of Chicago, Drexel University, Fermilab, the Institute for
  Advanced Study, the Japan Participation Group, Johns Hopkins University, the Joint Institute for Nuclear Astrophysics, the Kavli
  Institute for Particle Astrophysics and Cosmology, the Korean Scientist Group, the Chinese Academy of Sciences (LAMOST), Los
  Alamos National Laboratory, the Max-Planck-Institute for Astronomy (MPIA), the Max-Planck-Institute for Astrophysics (MPA), New
  Mexico State University, Ohio State University, University of Pittsburgh, University of Portsmouth, Princeton University, the
  United States Naval Observatory, and the University of Washington.}



\end{document}